# Diffusion pore imaging with generalized temporal gradient profiles

Frederik Bernd Laun[1,2], Tristan Anselm Kuder[1]

1) Medical Physics in Radiology, German Cancer Research Center (DKFZ), Im Neuenheimer Feld 280, 69120 Heidelberg, Germany

2) Quantitative Imaging-Based Disease Characterization, German Cancer Research Center (DKFZ), Im Neuenheimer Feld 280, 69120 Heidelberg, Germany

**Abstract**

In porous material research, one main interest of nuclear magnetic resonance diffusion (NMR) experiments is the determination of the shape of pores. While it has been a longstanding question if this is in principle achievable, it has been shown recently that it is indeed possible to perform NMR-based diffusion pore imaging. In this work we present a generalization of these previous results. We show that the specific temporal gradient profiles that were used so far are not unique as more general temporal diffusion gradient profiles may be used. These temporal gradient profiles may consist of any number of "short" gradient pulses, which fulfil the short-gradient approximation. Additionally, "long" gradient pulses of small amplitude may be present, which can be used to fulfil the rephasing condition for the complete profile. Some exceptions exist. For example, classical q-space gradients consisting of two short gradient pulses of opposite sign cannot be used as the phase information is lost due to the temporal antisymmetry of this profile.

## 1. Introduction

Nuclear magnetic resonance (NMR) based diffusion experiments find widespread application in medical imaging and porous media research as the measured diffusion-weighted signal allows one to infer information about structures restricting the diffusion process [1-3]. For a long time, it has been an open question if NMR diffusion experiments could exactly reveal the shape of the boundary of closed pores [4]. In a recent publication [5], we could show that this is indeed possible, if the temporal antisymmetry of the diffusion weighting magnetic field gradients is broken by using one long and one short diffusion weighting gradient pulse (gradient profile $G_{long-narrow}(t)$, see Fig. 1). In a porous sample with many closed pores of similar shape filled with an NMR visible medium, this technique can be applied to obtain an image of the pore geometry averaged over the considered volume element. Compared to conventional NMR imaging, largely increased signal-to-noise ratios can be achieved, since the signal can be collected from the whole sample. Shortly after publication of this first approach, Shemesh et al. [6] proposed another reconstruction approach that is based on multiple wave-vector diffusion weightings [7,8], for which Özarslan and Basser previously recognized that the signal could contain phase information about the pore shape [9]. Shemesh et al. [6] used a gradient profile consisting of three short gradient pulses ($G_{121}(t)$, see Fig. 1). Using their image reconstruction approach, this profile, in combination with the classical q-space profile [10,11] consisting of two short gradients ($G_{11}(t)$, see Fig. 1), can be used to infer the pore space function if one knows a priori that the pore space function is point symmetric [12]. We generalized these results and showed that the pore space function can be retrieved using the profile $G_{121}(t)$ without the need to assume point symmetry of the domain, and without the need to acquire additional data with the profile $G_{11}(t)$ [13].

It is a natural question to ask if these proposed gradient profiles have extraordinary properties, or if the main aspect is the broken temporal antisymmetry. In the latter case, one may ask which further gradient profiles are suitable to retrieve the shape of the pore space function. Answers to these questions may be helpful to asses, which acquisition scheme is best suited for a specific application. In this technical note, we show that a much larger class of temporal gradient profiles can be used as long as the profiles contain at least one short gradient pulse.

**2. Methods**

Assume that the gradient profile

$$\mathbf{G}(t) = \mathbf{G_0}(t) + \gamma^{-1}\mathbf{q}\sum_{n=1}^{N} c_n \delta(t-t_n) \tag{1}$$

is used, where $\gamma$ is the gyromagnetic ratio. $\mathbf{G_0}(t)$ is an arbitrary "small" gradient. A more thorough discussion of the term "small" and the small gradient can be found in [14] in section 4.1. In particular, $\mathbf{G_0}(t)$ should not contain Dirac delta functions and should not induce a considerable diffusion weighting effect on its own. The constants $c_n$ are real and one of them is equal to one, all others must be smaller than one and larger than minus one. The parameter $c_n\mathbf{q}$ specifies the wave-vector that is generated by the gradient pulse applied at time $t_n$, which is represented by the Dirac delta function $\delta(t-t_n)$. In general, the relation between a gradient pulse $\mathbf{G_n}(t)$ and its associated wave-vector is $\mathbf{q_n} = \gamma\int \mathbf{G_n}(t)dt$. The separation of the Dirac delta pulses must be "long", which means $|t_n - t_{n-1}| \gg L^2/D$ for all $t_n$, where $D$ is the free diffusion coefficient. $L$ is the length of the domain, for instance, the radius of a spherical pore. The upper bound $N$ is an arbitrary integer greater than or equal to one. As usual, the Dirac delta functions are considered as the limiting case of very strong and short diffusion gradients. It is further assumed that the rephasing condition is fulfilled:

$$\int_0^T \mathbf{G}(t)dt = 0 \tag{2}$$

Here, $T$ is the total duration of the gradient profile. We introduce the pore space function $\chi(\mathbf{x})$, which is one inside and zero outside the domain. $P(\mathbf{x_2}, \mathbf{x_1}, \Delta t)$ is the propagator indicating the probability for a particle to travel from $\mathbf{x_1}$ to $\mathbf{x_2}$ in the time $\Delta t$. First assume that the small gradient $\mathbf{G_0}(t)$ is zero. Then at least two short gradient pulses must be applied to fulfill the rephasing condition and the obtained signal, which is normalized on unity, is

$$S(\mathbf{q}) = \int d\mathbf{x_1} \frac{\chi(\mathbf{x_1})}{V} e^{-i\mathbf{q_1}\mathbf{x_1}} \prod_{n=2}^{N}\left(\int d\mathbf{x_n}\right) \prod_{n=2}^{N}\left(P(\mathbf{x_n}, \mathbf{x_{n-1}}, t_n - t_{n-1}) e^{-i\mathbf{q_n}\mathbf{x_n}}\right), \quad (3)$$

with $\mathbf{q_n} = \mathbf{q}c_n$. If $|t_n - t_{n-1}| \gg L^2/D$, the correlations of the positions $\mathbf{x_n}$ and $\mathbf{x_{n-1}}$ are lost and $P(\mathbf{x_n}, \mathbf{x_{n-1}}, t_n - t_{n-1})$ can be replaced by $\chi(\mathbf{x_n})/V$, where $V$ is the volume of the pore. Thus, Eq. (3) becomes

$$S(\mathbf{q}) = \frac{1}{V^N} \prod_{n=1}^{N} \int d\mathbf{x_n} \chi(\mathbf{x_n}) e^{-i\mathbf{q_n}\mathbf{x_n}}, \quad (4)$$

which is equal to

$$S(\mathbf{q}) = \prod_{n=1}^{N} \tilde{\chi}(\mathbf{q_n}) = \prod_{n=1}^{N} \tilde{\chi}(\mathbf{q}c_n), \quad (5)$$

where $\tilde{\chi}(\mathbf{q_n})$ is the Fourier transform of the pore space function $\chi(\mathbf{x_n})$:

$$\tilde{\chi}(\mathbf{q_n}) = \frac{1}{V} \int d\mathbf{x_n} \chi(\mathbf{x_n}) e^{-i\mathbf{x_n}\mathbf{q_n}} \quad (6)$$

Note that $\tilde{\chi}(-\mathbf{q_n}) = \tilde{\chi}^*(\mathbf{q_n})$, where the star denotes the complex conjugate, since $\chi(\mathbf{x_n})$ is real. If the small gradient is not zero, then, as laid out in [5,14], Eq. (5) must be modified with a phase factor $e^{-i\mathbf{q_0}\mathbf{x_{cm}}}$

$$S(\mathbf{q}) = e^{-i\mathbf{q_0}\mathbf{x_{cm}}} \prod_{n=1}^{N} \tilde{\chi}(\mathbf{q_n}) = e^{-i\mathbf{q_0}\mathbf{x_{cm}}} \prod_{n=1}^{N} \tilde{\chi}(c_n\mathbf{q}), \quad (7)$$

where $\mathbf{x_{cm}}$ is the center of mass of the domain, and $\mathbf{q_0}$ is

$$\mathbf{q_0} = \gamma \int_0^T \mathbf{G_0}(t) dt. \tag{8}$$

Due to the rephasing condition, identical pores at different locations yield identical signals; all pores seem to be shifted to the same location. Thus, the reconstructed pore image may be placed without loss of generality such that the center of mass of the pore is at the coordinate origin, such that $e^{-i\mathbf{q_0}\mathbf{x_{cm}}} = 1$.

The task is to solve Eq. (5) for $\tilde{\chi}(\mathbf{q})$. If $N$ was equal to one, $\tilde{\chi}(\mathbf{q})$ can be set equal to $S(\mathbf{q})$ and $\chi(\mathbf{x})$ can be calculated straightforwardly by Fourier transformation as described in [5,14]. If $N$ is larger than one, we assume that $m$ is the index for which $c_m = 1$. Thus, $\mathbf{q_m} = c_m \mathbf{q} = \mathbf{q}$ is the largest of the values $\mathbf{q_n}$ and we rewrite Eq. (5) as

$$\tilde{\chi}(\mathbf{q}) = S(\mathbf{q}) / \prod_{n=1, n \neq m}^{N} \tilde{\chi}(\mathbf{q_n}). \tag{9}$$

This is an equation that can be solved iteratively if $S(\mathbf{q})$ is known from experiments, and if $\tilde{\chi}(\mathbf{q_n})$ is known for some initial small $\mathbf{q_n}$-values. Considering Eq. (6), one finds $\tilde{\chi}(\mathbf{0}) = 1$. $\tilde{\chi}(\mathbf{q_n})$ can be expanded in a series for small q-values. As laid out in [13], one has freedom in specifying the term linear in $\mathbf{q_n}$ as this term specifies the position of the reconstructed pore. If this linear term is set to zero, the pore is reconstructed such that its center of mass is located at the origin. By these considerations, $\tilde{\chi}(\mathbf{q_n})$ is fully specified for small q-values, and the iterative solution of Eq. (9) is possible and unique, for example by setting $\tilde{\chi}(\mathbf{q_n}) = 1$ for the initial small values of $\mathbf{q_n}$.

For demonstration purposes, we chose the four temporal gradient profiles $G_{12}(t)$, $G_{123}(t)$, $G_{235}(t)$ and $G_{1113}(t)$ (see Fig. 1, $\delta$ is the duration of the short gradient pulse). Simulations based on the multiple correlation function technique [4,14-19] were performed using the hemi-equilateral triangle and the spherical domain. The following computation parameters were chosen: Free diffusion coefficient $D$ =1 µm²/ms, length of the longest edge of the triangle $L$ =20 µm, the radius of the sphere is also denoted by $L$ and is set to 10 µm, $T$ =1 s, duration of the short gradient pulses 1 ms, 100 eigenvalues. The maximal q-value was 4 µm$^{-1}$ for the equilateral triangle and 2 µm$^{-1}$ for the sphere. Data points were acquired radially at 400 equidistant q-values for the equilateral triangle and at 300 equidistant q-values for the sphere. Simulations with three parameter setting were performed:

(i) To compute diffusion pore images, the q-space was sampled radially along 120 isotropically distributed spokes. The image was reconstructed by inverse Radon transform for a field of view of 31.4 µm.

(ii) Simulations identical to (i) were performed, except for the addition of Gaussian noise with standard deviation $\sigma = 1/1000$ on real and imaginary part of the signal. This corresponds to a signal-to-noise ratio (SNR) of 1000 as the signal is normalized on unity.

(iii) To assess the statistics of the reconstructed $\tilde{\chi}(\mathbf{q})$, the spoke along the x-direction was simulated $10^5$ times to estimate the statistics of the reconstructed $\tilde{\chi}(\mathbf{q})$. The median, lower and upper quartiles of $\tilde{\chi}(\mathbf{q})$ were calculated. These simulations were performed with $\sigma = 1/100$ and $\sigma = 1/1000$.

In all simulations, Eq. (9) was solved iteratively. The first value of $\tilde{\chi}(\mathbf{q})$ was set to one. The second was estimated using the fact that $\tilde{\chi}(\mathbf{q}) \approx 1 - a\mathbf{q}^2$ and $S(\mathbf{q}) \approx 1 - b\mathbf{q}^2$ for small q-values with some constants $a$ and $b$ [13]. This leads to the condition

$$1 - b\mathbf{q}^2 = \prod_n^N \left(1 - ac_n^2\mathbf{q}^2\right) = 1 - a\mathbf{q}^2 \sum_n^N c_n^2 + O\left(\mathbf{q}^3\right) \tag{9}$$

and thus one finds

$$a = b / \sum_n^N c_n^2. \tag{9}$$

The second value of $\tilde{\chi}(\mathbf{q})$ was set according to Eq. (11) and a basic iteration scheme was used (for details on the implementation see supplemental material, where the Matlab-code is provided). For the sphere, a basic outlier rejection was implemented, setting values of $\tilde{\chi}(\mathbf{q})$ for which $|\tilde{\chi}(\mathbf{q})| > 1$ to zero. This was necessary since $\tilde{\chi}(\mathbf{q})$ of the sphere has more zero crossings (which is problematic, see Eq. (9)).

**3. Results**

Figure 2 shows the influence of noise on the reconstruction. Black and gray lines represent real and imaginary part of the reconstructed $\tilde{\chi}(\mathbf{q})$. The three lines of each color represent the lower quartile, the median and the upper quartile. The deviation of the quartiles from the median at small q-values is minimal but it increases continuously due to the iterative reconstruction. This shows that low-resolution images can be acquired much easier than high-resolution images. At SNR=1000, the reconstruction is the more stable, the fewer short gradient pulses are used: For instance, the reconstruction with 12-gradients is more stable than that with 121-gradients, which is still more stable than the reconstruction with 1113-gradients. At SNR=100, this picture changes slightly and becomes less clear-cut. For example, the reconstruction with 1113-gradients is more stable than with 121-gradients at $qL = 30$, but it is less stable at $qL = 40$. The long gradient that is present in the profiles $G_{123}(t)$ and $G_{12}(t)$ leads to a blurring of the image and thus to a decreased amplitude of $\tilde{\chi}(\mathbf{q})$ at large $qL$.

Figure 3 shows similar graphs as Fig. 2, but for the spherical domain. The spherical domain has two properties which make the reconstruction more complicated than for the hemi-equilateral triangle.

Firstly, the function $\tilde{\chi}(\mathbf{q})$ has more zero-crossings. This can be clearly appreciated in Fig. 3. The reconstruction often fails at and beyond the zero crossing. The reconstruction algorithm that was used here is quite basic and obstinately uses Eq. (9) for the iterative reconstruction (see supplemental material). More sophisticated algorithms might attenuate this problem. At large q-values, the reconstructed $\tilde{\chi}(\mathbf{q})$ is zero in Fig. 4 due to the applied outlier rejection.

Secondly, $\tilde{\chi}(\mathbf{q})$ drops quicker with increasing q-value for the sphere than for the hemi-equilateral triangle (see appendix A). The signal (see Eq. (5)) does indeed drop so quickly, that increasing the SNR does hardly increase the accessible region in q-space (for a more detailed discussion of the influence of noise depending on the number of applied short gradient pulses see appendix A). This is observable in Fig. 3: The first zero crossing of $\tilde{\chi}(\mathbf{q})$ can be resolved with SNR=100 and SNR=1000. Values beyond the second zero crossing of $\tilde{\chi}(\mathbf{q})$ are not accessible with any of the gradient profiles, except for, remarkably, by the 1113-gradients (at least to some extent). The advantage of the 1113-gradients in this regard is, that this gradient profile samples $\tilde{\chi}(\mathbf{q})$ thrice at small q-values and only once at a large q-value. This is advantageous if $\tilde{\chi}(\mathbf{q})$ drops quickly at larger q-values.

Figures 4 and 5 show the reconstructed images. The triangular and spherical shapes are clearly visible (for details see figure caption).

## 4. Discussion

It turns out that the pore space function can be detected with generalized gradient profiles using the iterative approach presented here, excluding two important cases. First, at least one short gradient pulse must be present, because, otherwise, no $\tilde{\chi}(\mathbf{q_n})$-term appears in Eq. (5). Second, the iterative reconstruction of Eq. (9) is not possible if all short gradients generate the same wave vector, or wave vectors of opposite sign. Thus, classical q-space imaging gradients are not suitable for this approach.

The findings presented here are foremost of theoretical interest, but some comments regarding practical implementations should be made. First, it should be noted that the iterative reconstruction of Eq. (9) is crucially depending on the exact cancellation of the signal $S(\mathbf{q})$ and of $\tilde{\chi}(\mathbf{q_n})$ at the zero crossings of $\tilde{\chi}(\mathbf{q_n})$. This is not a problem under ideal conditions, since $S(\mathbf{q})=0$ if there is a value for $n$ for which $\tilde{\chi}(\mathbf{q_n})=0$. Measurements, however, are not ideal in several regards. For example, field inhomogeneities and measurement noise are present and, moreover, the actual $S(\mathbf{q})$ deviates from the theoretical $S(\mathbf{q})$, since the total duration $T$ is not infinite. The hemi-equilateral triangle is a benign domain in this regard, since zero crossings of $\tilde{\chi}(\mathbf{q_n})$ are sparse, but, for instance, cylindrical and spherical domains are much more prone to "no-cancellation artifacts" as $\tilde{\chi}(\mathbf{q_n})$ exhibits many zero crossings. This can be circumvented by straightforward approaches such as, for example, assuming the actual pore shape is known a priori, which allows a direct fit as described in [6]. Otherwise, it may be necessary to implement more sophisticated reconstruction algorithms.

What is the ideal number of short gradient pulses in real experiments? In our opinion, there are basically two options. The first option is to use only one short gradient pulse. This is beneficial since the iterative reconstruction using Eq. (9) is not required, and since a Fourier transform can be used to retrieve the pore image straightforwardly. Using only one short gradient comes at two big costs. First, a long small gradient must be used, which rules out attempts to use stimulated echoes (e.g. [20-22]), and thus reduces the available sequence options. Second, the convergence towards the long-time limit, in which a sharp image can be reconstructed, is slower than with short gradient pulses only. Details on convergence properties are described in [14].

Both these complications can be avoided using solely narrow gradient pulses and the iterative reconstruction, which, however, is less stable than the one-pulse Fourier transform [13]. In general, it seems beneficial to use as few short gradient pulses as possible. More short gradient pulses require a longer diffusion time and thus entail a reduced signal owing to relaxation. Moreover, more radio-frequency pulses must be used if the magnetization shall be stored along the longitudinal direction between the short gradient pulses. This makes sequences more complicated and reduces the SNR [23]. Nonetheless, using more than three gradient pulses might be beneficial under some special circumstances, for example, if the signal drops quickly at larger q-values as for the spherical domain (Fig. 3), or to avoid zero-crossings. To minimize the effect of zero crossings, it might also be helpful to combine data acquired with different gradient profiles.

A further point must be noted when discussing whether long-narrow gradients or the other gradient profiles should be used. The most appealing property of diffusion pore imaging is that the signal of the whole sample is measured and one can thus hope to acquire an image of an average pore shape. As the pores in a real sample are usually not identical, one important demand on the technique is that it should be applicable to ensembles of differently shaped pores. In the discussion of our previous work [13], we erroneously conjectured that for both long-narrow gradients and 121-gradients such an average pore could be measured in general. As shown in Appendix B, this statement is only valid for the long-narrow gradient profile. In any case, the signal is averaged over all pores. However, in the reconstructed pore image, of all gradient profiles we have tested so far, only the long-narrow gradient profile and derivations of it directly yield the arithmetic average of all pore space functions, the others don't. A detailed investigation under which circumstances the other gradient profiles can nonetheless be used in the presence of size or shape distributions is beyond the scope of this article.

In conclusion, we have shown that the shape of arbitrary pores can be retrieved with quite general diffusion gradient profiles. The profiles presented in previous works [5,6,13,14,24] are thus not special, and one has freedom in shaping the gradient profile to match experimental conditions. Three questions remain open. First, is there a gradient profile without short gradient pulse that can be used to retrieve the pore space function? Such an approach would presumably reduce the demands on the gradient system. Second, how much information can be gained in open and permeable systems? Third, in how far is it possible to one image ensembles of differently shaped pores with the proposed generalized gradient profiles. Answers to these questions would be highly valuable.

*Appendix A – Influence of Noise*

We consider the case without long gradients. Then, according to Eq. (5), the signal is given by $S(\mathbf{q}) = \prod_{n=1}^{N} \tilde{\chi}(\mathbf{q_n})$. Some examples of $\tilde{\chi}(\mathbf{q})$ for basic domains are

$$\tilde{\chi}_{slab}(q) = \frac{2\sin(qL/2)}{qL} \tag{9}$$

$$\tilde{\chi}_{sphere}(\mathbf{q}) = 3(qL)^{-3}\left(-qL\cos(qL) + \sin(qL)\right) \tag{9}$$

$$\tilde{\chi}_{hemi}\left((q,0,0)^{\mathrm{T}}\right) = (qL)^{-2}\left(8 - 8e^{-iqL/2} - 4iqL\right)e^{iqL/6} \tag{9}$$

with $q = |\mathbf{q}|$. For the slab domain, $L$ is the distance of the confining plates. Neglecting oscillations, one finds at large $q$

$$\tilde{\chi}_{slab}(q) \propto \frac{2}{qL} \tag{9}$$

$$\tilde{\chi}_{sphere}(\mathbf{q}) \propto \frac{3}{(qL)^2} \tag{9}$$

$$\tilde{\chi}_{hemi}\left((q,0,0)^{\mathrm{T}}\right) \propto \frac{4}{qL} \tag{9}$$

Thus $\tilde{\chi}(\mathbf{q})$ often drops with $b_1\mathbf{q}^{-n}$ at large $q$ where $n \geq 1$, and where $b_1$ is a domain-dependent constant. Using the approximation given by Eqs. (15) to (17), one finds for the signals $S_{long-narrow}$, $S_{12}$, $S_{121}$ and $S_{1113}$ that are acquired with $G_{long-narrow}(t)$, $G_{12}(t)$, $G_{121}(t)$ and $G_{1113}(t)$

$$S_{long-narrow} \propto b_1 \mathbf{q}^{-n} \tag{9}$$

$$S_{12} \propto b_1(0.5\mathbf{q})^{-n} b_1\mathbf{q}^{-n} \tag{9}$$

$$S_{121} \propto b_1(0.5\mathbf{q})^{-n} b_1 \mathbf{q}^{-n} b_1(0.5\mathbf{q})^{-n} \tag{9}$$

$$S_{1113} \propto b_1(0.33\mathbf{q})^{-n} b_1(0.33\mathbf{q})^{-n} b_1(0.33\mathbf{q})^{-n} b_1 \mathbf{q}^{-n} \tag{9}$$

If the signal is smaller than the noise level, the reconstruction will fail. Thus $q$ should not exceed a maximal q-value $q_{max}$, for which $S = \sigma = 1/SNR$. Thus one finds:

- Long-narrow gradients: $q_{max} = (b_1 SNR)^{1/n}$

- 12-gradients: $q_{max} = 2^{1/2} (b_1^2 SNR)^{1/2n}$

- 121-gradients: $q_{max} = 4^{1/3} (b_1^3 SNR)^{1/3n}$

- 1113-gradients: $q_{max} = 27^{1/4} (b_1^4 SNR)^{1/4n}$

- General gradients with $c_1, c_2, \ldots c_N$: $q_{max} = \left(\prod_i^N c_i^{-1}\right)^{1/N} (b_1^N SNR)^{1/Nn}$

There are two competing factors. Take the 1113-gradients: Many small gradients generate a large factor inside the brackets (the number 27 in this example), which is favorable. This results from the fact that short gradients with small q-values sample the $\tilde{\chi}(\mathbf{q})$-function at smaller q-values, where $\tilde{\chi}(\mathbf{q})$ is large. On the other hand, the exponent $1/4n$ decreases $q_{max}$. This factor can be attributed to the effect that each additional short gradient pulse results in an additional $\tilde{\chi}(\mathbf{q})$-term in Eq. (5).

TABLE A1. $L \cdot q_{max}$ for hemi-equilateral triangle and sphere.

| domain | hemi-equilateral triangle | | sphere | |
|---|---|---|---|---|
| SNR | 100 | 1000 | 100 | 10000 |
| $n$ | 1 | 1 | 2 | 2 |
| $b_1$ | 4 | 4 | 3 | 3 |
| $L \cdot q_{max}$, long-narrow | 400 | 4000 | 17.3 | 54.8 |
| $L \cdot q_{max}$, 12-gradients | 56.6 | 179 | 7.75 | 13.8 |
| $L \cdot q_{max}$, 121-gradients | 29.5 | 63.5 | 5.92 | 8.69 |
| $L \cdot q_{max}$, 1113-gradients | 28.8 | 51.3 | 7.02 | 9.36 |

Table A1 shows numerical examples of $L \cdot q_{max}$ for the hemi-equilateral triangle measured along the x-direction and for the sphere. The two SNRs of 100 and 1000 are considered. For the long-narrow gradient profile, increasing SNR increases $q_{max}$ considerably. Using more short gradient pulses, it becomes increasingly difficult to increase $q_{max}$ by increasing the SNR, especially for the sphere. For low SNR and larger $n$, $q_{max}$ is larger when using 1113-gradients instead of 121-gradients. A natural question arising at this point is: Does increasing the number of gradient pulses further increase this $q_{max}$-value and thus the SNR efficiency?

Fig. A1 shows a plot similar to Fig. 3, simulated with the gradient profile $G_{1111111119}(t)$, which consists of nine short gradient pulses generating q-values $-q/9$ and one short gradient pulse generating a q-value $q$. Up to $qL = 60$, the reconstruction is even more stable than that with $G_{121}(t)$ (compare to Fig. 2b).

If this approach it pushed further, such that the number of small short gradient pulses is not only nine, but a large number $M$, one finds for the signal

$$S(\mathbf{q}) = \tilde{\chi}^M(\mathbf{q}/M)\tilde{\chi}(\mathbf{q}) \tag{9}$$

At small q-values, $\tilde{\chi}(\mathbf{q}) \approx 1 - a\mathbf{q}^2$ holds true, with the domain-dependent constant $a$. Thus Eq. (22) becomes

$$S(\mathbf{q}) \approx \left(1 - a\frac{q^2}{M^2}\right)^M \cdot \tilde{\chi}(\mathbf{q}) \approx \left(1 - Ma\frac{q^2}{M^2}\right)\tilde{\chi}(\mathbf{q}) \approx \tilde{\chi}(\mathbf{q}) \tag{9}$$

In the limit of large $M$ (and ignoring relaxation), this profile, which we call $G_{111...M}(t)$ becomes as SNR-efficient as the long-narrow gradient profile! This is astonishing, but the analogy goes even further. Consider, for example, the case of a pore shifted by $\mathbf{x}_{shift}$, for which the shifted $\tilde{\chi}(\mathbf{q})$ becomes

$$\tilde{\chi}_{shift}(\mathbf{q}) = \tilde{\chi}(\mathbf{q})e^{-i\mathbf{q}\cdot\mathbf{x}_{shift}} \tag{9}$$

Then, the contribution of the small gradients is

$$\tilde{\chi}_{shift}^{M}(\mathbf{q}/M) = \tilde{\chi}^{M}(\mathbf{q}/M)\left(e^{-i\mathbf{q}\cdot\mathbf{x}_{shift}/M}\right)^{M} \approx 1 \cdot e^{-i\mathbf{q}\cdot\mathbf{x}_{shift}} = e^{-i\mathbf{q}\cdot\mathbf{x}_{shift}} \tag{9}$$

The gradient profile $G_{111...M}(t)$ is not only as SNR-efficient as $G_{long-narrow}(t)$, it is indeed identical regarding accumulated phase of a diffusing particle! Both have "small" gradients which are distributed over a long time interval generating a phase proportional to the center of mass of the domain and one short gradient pulse generating the final q-value.

*Appendix B – Ensembles of differently shaped pores*

Suppose that two differently shaped pores of equal volume, which are described by $\tilde{\chi}_1(\mathbf{q})$ and $\tilde{\chi}_2(\mathbf{q})$, are measured with the gradient profile $G_{121}(t)$. The resulting signal of these separated pores is

$$S_s(\mathbf{q}) = \frac{1}{2}\tilde{\chi}_1(\mathbf{q})\tilde{\chi}_1^2(-\mathbf{q}/2) + \frac{1}{2}\tilde{\chi}_2(\mathbf{q})\tilde{\chi}_2^2(-\mathbf{q}/2) \tag{9}$$

The average image is described by $\tilde{\chi}_a(\mathbf{q}) = (\tilde{\chi}_1(\mathbf{q}) + \tilde{\chi}_2(\mathbf{q}))/2$ for which one would measure

$$\begin{aligned}S_a(\mathbf{q}) &= \frac{1}{8}\tilde{\chi}_a(\mathbf{q})\tilde{\chi}_a^2(-\mathbf{q}/2) \\ &= \frac{1}{8}\tilde{\chi}_1(\mathbf{q})\tilde{\chi}_1^2(-\mathbf{q}/2) + \frac{1}{8}\tilde{\chi}_2(\mathbf{q})\tilde{\chi}_2^2(-\mathbf{q}/2) + \text{other terms} \\ &\neq S_s(\mathbf{q})\end{aligned} \tag{9}$$

Since $S_s(\mathbf{q})$ and $S_a(\mathbf{q})$ are not identical, one cannot straightforwardly measure an average image using the gradient profile $G_{121}(t)$ in general. In this context, the same argument, that "the average of the product is not equal to the product of the averages", has been raised by Kiselev and Novikov in their recent Comment [12].

For $G_{long-narrow}(t)$, the situation is different. The signal for separated and "averaged" pores is given by

$$S_{long-narrow}(q) = \frac{1}{2}\tilde{\chi}_1(\mathbf{q}) + \frac{1}{2}\tilde{\chi}_2(\mathbf{q}).$$

Thus, the Fourier transform directly yields an average pore image.

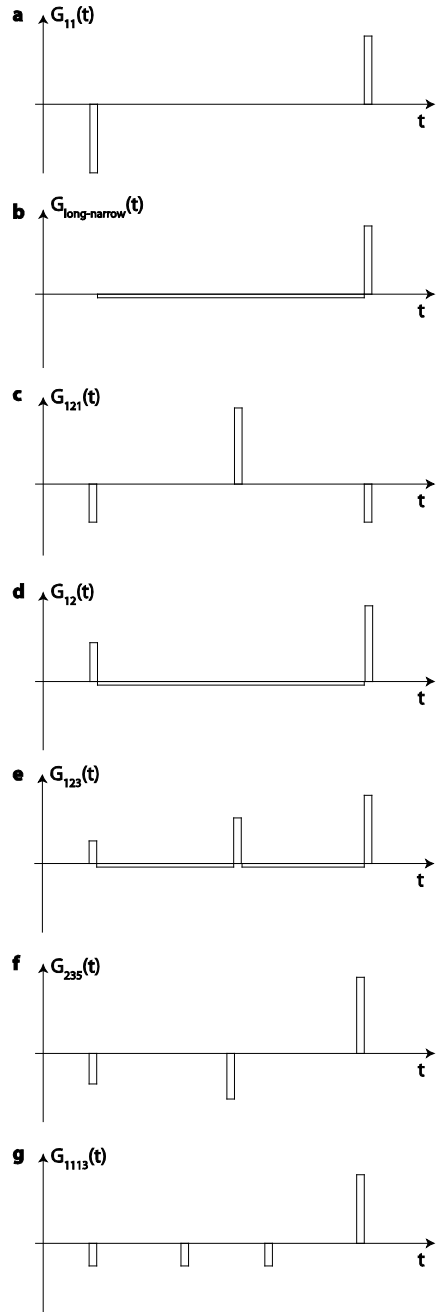

**Fig. 1.** The gradient profiles used in this manuscript. a) Classical q-space gradients $G_{11}(t)$. b) "Long-narrow" gradients $G_{long-narrow}(t)$ consisting of one long and one narrow gradient pulse. The zeroth moment of both gradient pulses cancel in order to fulfill the rephasing condition. c) Gradient profile $G_{121}(t)$ consisting of three narrow gradient pulses. d) Gradient profile $G_{12}(t)$ consisting of two narrow and one long gradient pulse. e) Gradient profile $G_{123}(t)$. The relative zeroth moment of the narrow gradient pulses is 1/2/3. f) Gradient profile $G_{235}(t)$ consisting of three narrow gradient pulses with relative zeroth moments of -2/-3/5. g) Gradient profile $G_{1113}(t)$ consisting of four narrow gradient pulses with relative zeroth moments of -1/-1/-1/3. In the text, the gradient profiles will also be labeled as, e.g., 1113-gradients.

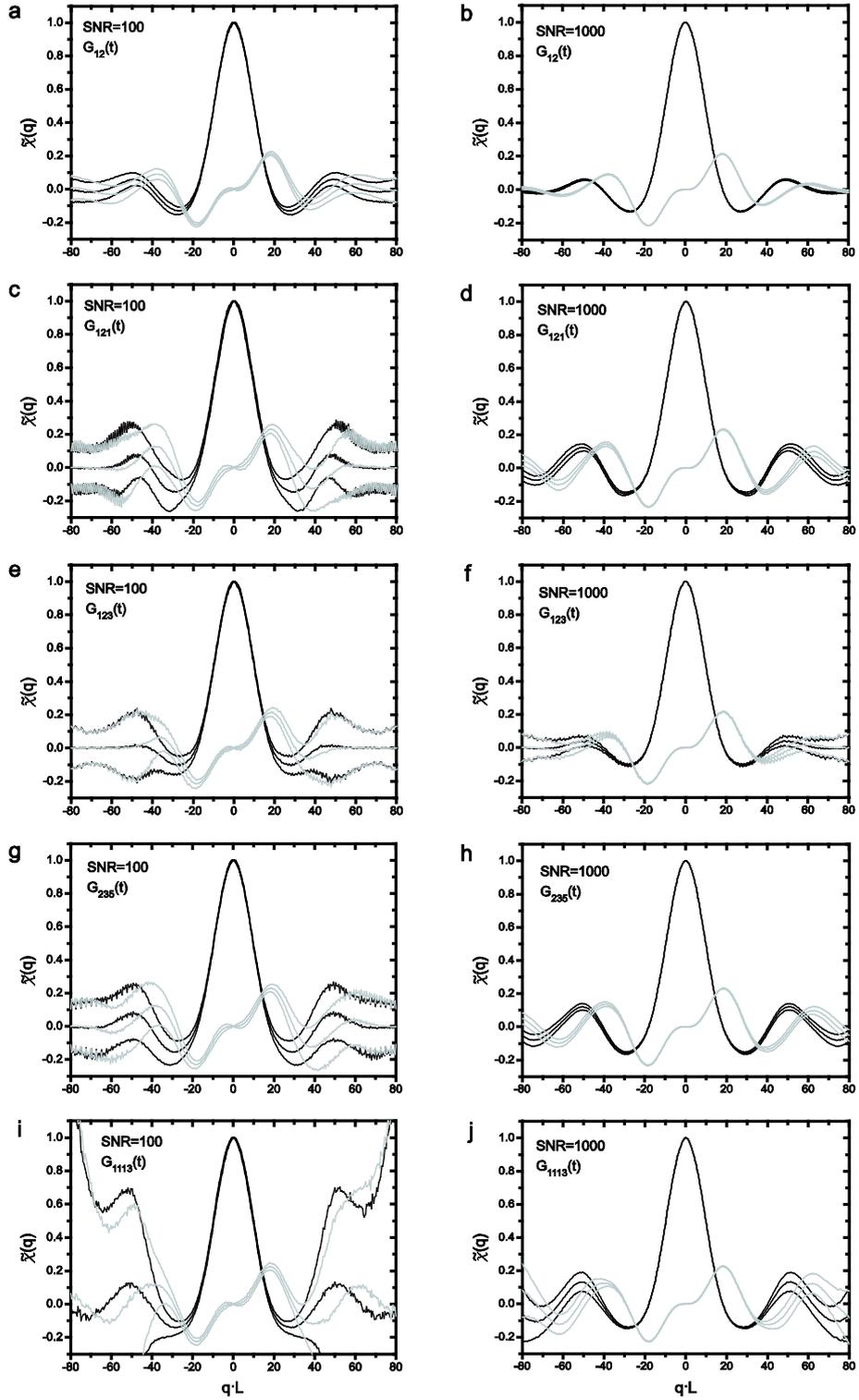

**Fig. 2.** Reconstructed $\tilde{\chi}(\mathbf{q})$ of the hemi-equilateral triangle with the diffusion gradient pointing along the x-direction. Black lines represent the real part and gray lines represent the imaginary part of $\tilde{\chi}(\mathbf{q})$. The three lines of each color represent the lower quartile, the median and the upper quartile of 10000 reconstructed $\tilde{\chi}(\mathbf{q})$.

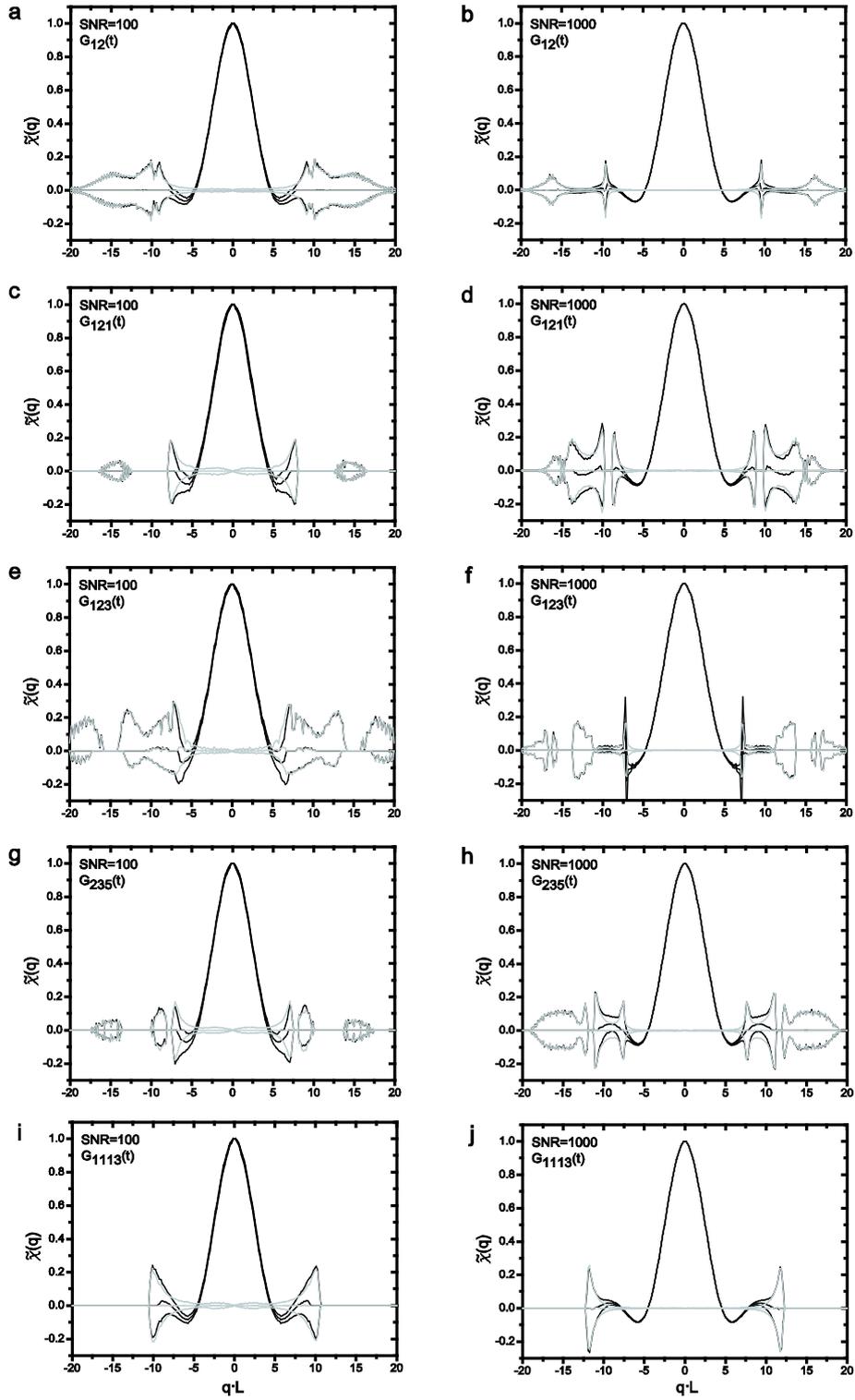

**Fig. 3.** Reconstructed $\tilde{\chi}(\mathbf{q})$ of the spherical domain with the diffusion gradient pointing along the x-direction. Black lines represent the real part and gray lines represent the imaginary part of $\tilde{\chi}(\mathbf{q})$. The three lines of each color represent the lower quartile, the median and the upper quartile of 10000 reconstructed $\tilde{\chi}(\mathbf{q})$.

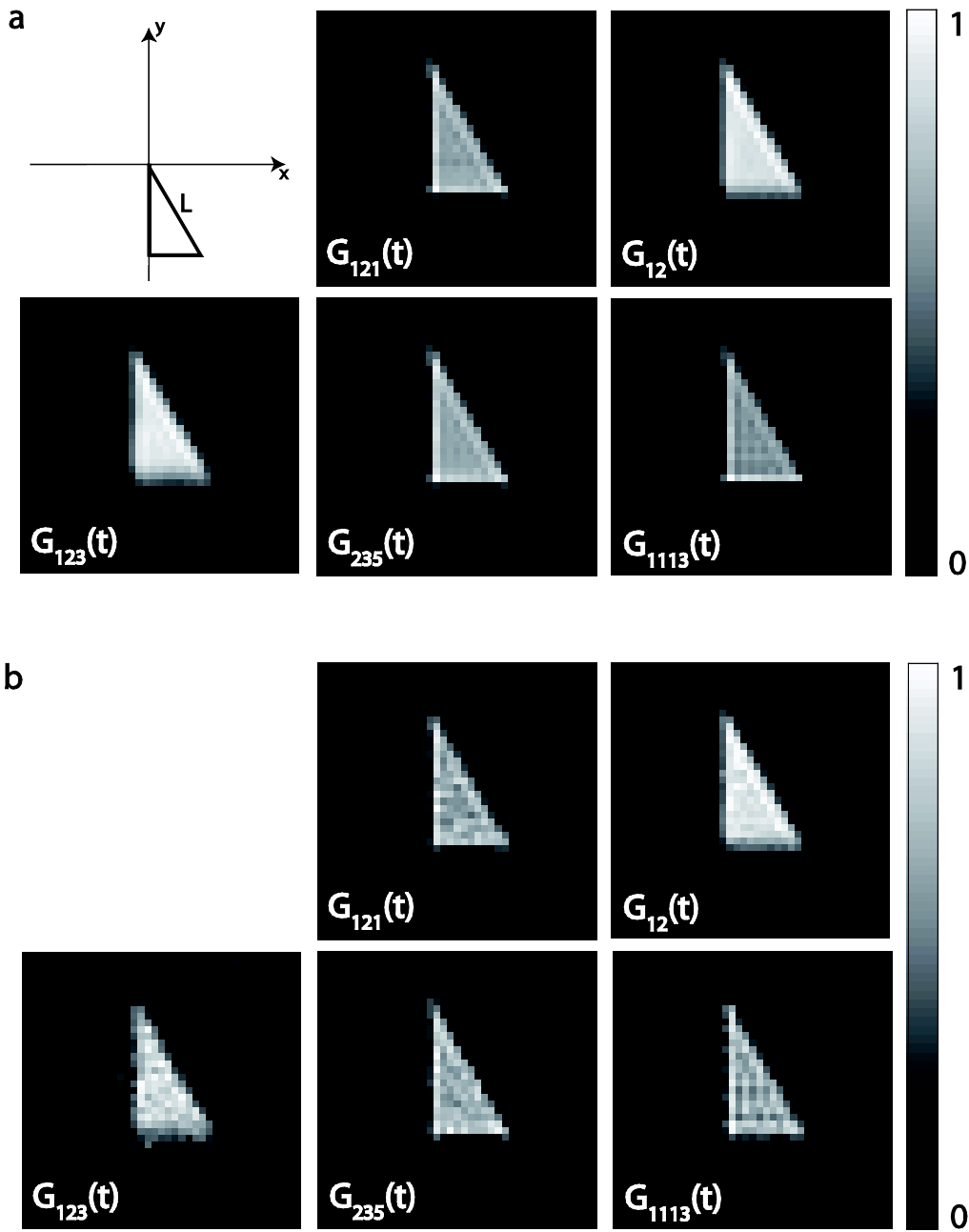

**Fig. 4.** Diffusion pore images obtained with a variety of gradient profiles for the hemi-equilateral triangle depicted in the upper left sketch. Images were min-max normalized. a) SNR is infinite. Unlike the other gradient profiles, the profiles $G_{12}(t)$ and $G_{123}(t)$ have a long gradient between the short gradient pulses, which leads to a blurring of the image. This blurring reduces at longer diffusion times (data not shown). The finite duration of the short gradient pulses leads to an edge enhancement effect. The images obtained with $G_{12}(t)$ and $G_{123}(t)$ appear brighter because the images were min-max normalized and because the edge-enhancement, which especially prominent in the corners, is smeared out. b) SNR = 1000. The image obtained with $G_{12}(t)$ is less noisy than those obtained with the other gradient profiles, which are approximately of equal quality for this example.

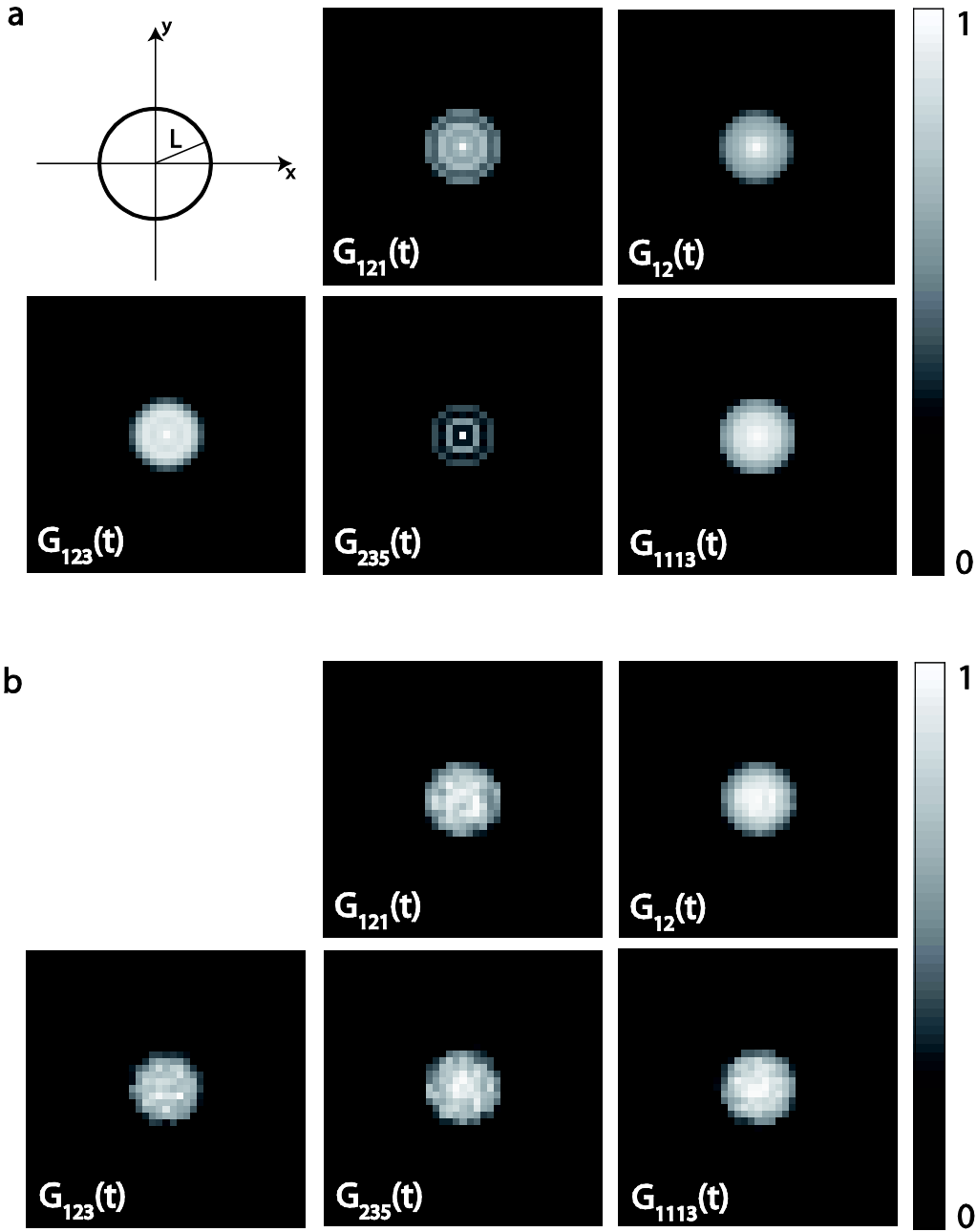

**Fig. 5.** Diffusion pore images obtained for the spherical domain depicted in the upper left sketch with a variety of gradient profiles. Images were min-max normalized. The reconstruction was performed in two dimensions, such that the image should be of higher intensity in the center than at the rim showing a projection of the spherical pore onto a 2-dimensional plane. a) SNR is infinite. For $G_{121}(t)$ and $G_{235}(t)$, pronounced oscillations are visible, that were not observed for the hemi-equilateral triangle, which is a more benign domain with less zero crossings of $\tilde{\chi}(\mathbf{q})$. The blurring generated by the long gradients of the gradient profiles $G_{12}(t)$ and $G_{123}(t)$ seems to reduce these oscillations. The short gradient pulses of smaller amplitude of the gradient profile $G_{1113}(t)$ sample most of the $\tilde{\chi}(\mathbf{q})$-curve before the first zero crossing in this example, which also reduces the oscillation artifact. b) SNR=1000. $G_{12}(t)$ yields the most stable image. The noise remarkably reduces the oscillations observed in (a) as it results in fluctuations of the values affected by the numerical instabilities near the zero crossings.

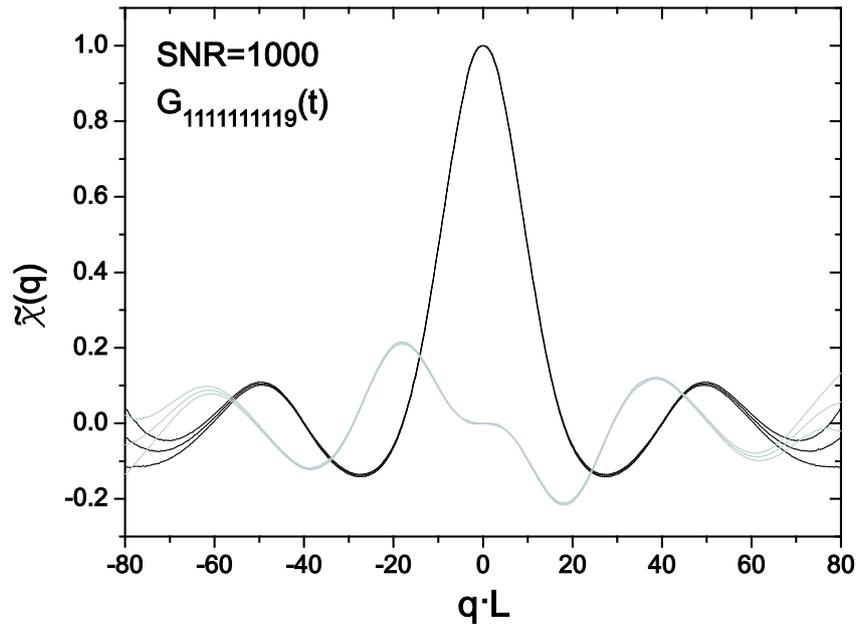

**Fig. A1.** Reconstructed $\tilde{\chi}(\mathbf{q})$ of the hemi-equilateral triangle with the diffusion gradient pointing along the x-direction. Black lines represent the real part and gray lines represent the imaginary part of $\tilde{\chi}(\mathbf{q})$. The three lines of each color represent the lower quartile, the median and the upper quartile of 10000 reconstructed $\tilde{\chi}(\mathbf{q})$.